\documentclass[10pt,letterpaper,twocolumn]{article} 

\usepackage{ol2}
\usepackage[draft]{hyperref}
\usepackage{amsmath}

\begin{document}

\twocolumn[ 

\title{Superradiance Paradox in waveguide lattices}


\author{Stefano Longhi}
\address{Dipartimento di Fisica, Politecnico di Milano and Istituto di Fotonica e Nanotecnologie del Consiglio Nazionale delle Ricerche, Piazza L. da Vinci 32, I-20133 Milano, Italy (stefano.longhi@polimi.it)}
\address{IFISC (UIB-CSIC), Instituto de Fisica Interdisciplinar y Sistemas Complejos, E-07122 Palma de Mallorca, Spain}

\begin{abstract}
Recently, it has been suggested that the collective radiative decay of  two point-like quantum emitters coupled to a waveguide,
separated by a distance comparable to the
coherence length of a spontaneously emitted photon, 
leads to an apparent $^{\prime}$superradiance paradox$^{\prime}$
by which one cannot decide whether independent or 
collective emission occurs. The resolution of the paradox stems from the strong non-Markovian dynamics arising from 
 the delayed field-mediated atom interaction. Here we suggest an integrated optics platform to emulate the superradiance 
 paradox, based on photon escape dynamics in waveguide lattices. Remarkably, Markovian decay dynamics and independent photon emission 
 can be restored by frequent (Zeno-like) observation of the system.
\end{abstract}

 ] 

{\it Introduction.}  Superradiance, introduced by Dicke in a pioneering paper \cite{r1}, is a fundamental optical phenomenon of collective spontaneous
emission of atoms \cite{r2,r3}. The simplest case of is the one of two identical atoms located at a distance within the photon wavelength. In this situation, light emitted by one atom may be reabsorbed by the other one and cooperative processes in the emission can arise. Specifically, if the atoms are initially in an entangled state where the excitation is distributed between both emitters, two possibilities, referred to as superradiance and subradiance regimes, can be observed: the system will either decay at twice the single atom
rate $\gamma$, or it will not radiate. Since the theoretical prediction of super- and sub-radiance states, a number of experiments have been designed to observe cooperative emission between two 
quantum emitters \cite{r5,r6,r7,r8}. When the two emitters are macroscopically separated by a distance of the order of the coherent length of the emitted photon, cooperative spontaneous emission is deeply influenced by retardation effects \cite{r9,r10}, exhibiting strong non-Markovian features \cite{r11}. Very recently, it has been argued that a seemingly paradoxical situation (referred to as the $^{\prime}$superradiance paradoxic$^{\prime}$ \cite{r11}) arises for macroscopically-separated quantum emitters, by which one cannot decide whether spontaneous emission occurs cooperatively or independently. The paradox is solved because of non-Markovian dynamics arising from coherent time-delayed feedback, indicating that spontaneous decay can be enhanced beyond the usual Dicke value  $2 \gamma$. Possible experimental quantum platforms to observe the superradiance paradox have been proposed, including quantum dots in photonic waveguides, atoms near optical nanofibers, and superconducting qubits side-coupled to waveguides \cite{r11}. 
While usually formulated in quantum optics context \cite{r2,r3}, superradiance is rooted in the interference of coherently radiating emitters, and thus 
has many classical aspects \cite{r4b} which can be observed in classical models as well \cite{r4}.\par 
In this Letter we suggest an integrated optic waveguide platform \cite{r12,r12b,r13,r14,r15,r16,r17,r18,r19,r20,r21,r22,r23} in which the escape dynamics of light waves in two waveguides, side-coupled to a waveguide lattice, shows cooperative effects like in spontaneous emission of two macroscopically-separated quantum emitters on a waveguide. The results illustrate a feasible route to probing the superradiance paradox in an entirely classical platform. Furthermore, the integrated optic waveguide setup can be feasible to test suppression of cooperative emission by frequent (Zeno-like) measurements of the system, that would be hard to observe in quantum electrodynamics settings.\\
\par
{\it The superradiance paradox.} Let us briefly remind the basic model of spontaneous emission of two macroscopically-spaced point-like atoms coupled to a non-chiral waveguide and the superradiance paradox \cite{r11}. The schematic of the system is shown in Fig.1(a). The two atoms are spaced by a distance $d$ such that $l_c/2<d< l_c$, where $l_c= v_g/ \gamma$ is the coherence length of the photon emitted by each atom individually, $v_g$ is the group velocity of the field and $\gamma$ the spontaneous decay rate of each individual emitter. The paradox can be stated as follows \cite{r11}: if we assume a collective spontaneous emission with the atoms initially prepared in a superradiant state, the spontaneous emission rate would be $2 \gamma$, so as the coherence length of the cooperatively-emitted photon would be $l_c/2$, i.e. smaller  than the separation $d$ of the two atoms. But in this case one could distinguish which atoms has emitted the photon, so we would exclude superradiance. On the other hand, if we assume the atoms to emit independently, the coherence length of the emitted photon would be $l_c$, i.e. longer than the atom spacing $d$. But in this case interference between decaying atomic channels is expected, i.e. one would expect cooperative spontaneous emission. Such a seemingly paradoxical situation is resolved by considering that, when the atomic spacing $d$ is of the order of $l_c$, the decay dynamics acquires strong non-Markovian features because of the retarded atom-atom interaction mediated by the photon field \cite{r11}. To study non-Markovian dynamics, let us consider the Hamiltonian of the entire system $\hat{H}=\hat{H}_A+\hat{H}_F+\hat{H}_{int}$ where
\begin{equation}
\hat{H}_A=\omega_0 \left( \hat{\sigma}^{(1)}_{+} \hat{\sigma}^{(1)}_{-}+\hat{\sigma}^{(2)}_{+}\hat{\sigma}^{(2)}_{-} \right), \; \hat{H}_F= \int dk \omega(k) \hat{a}^{\dag}(k) \hat{a}(k)
\end{equation}  
are the free Hamiltonians of the two identical atoms  and of the photon field in the waveguide and
\begin{equation}
\hat{H}_{int}= \int dk \left( g(k) \hat{\sigma}_{+}^{(1) }\hat{a}(k)+g(k) \exp(ikd) \hat{\sigma}_{+}^{(2)} \hat{a}(k) +H.c.   \right)
\end{equation} 
is the interaction Hamiltonian in the electric-dipole and rotating-wave approximation. In Eqs.(1) and (2), $\sigma_{\pm}^{(1,2)}$ are the raising/lowering atomic operators between ground $|g \rangle$ and excited $|e \rangle$ levels of the two emitters, $\omega_0$ is the atomic resonance frequency, $\hat{a}(k), \hat{a}^{\dag}(k)$ are the destruction/creation operators of the photon field in the waveguide with wave number $k$ ($k>0$ for progressive waves, $k<0$ for regressive waves), $\omega(k)$ is the dispersion relation of the waveguide ($ \omega(k) \simeq v_g |k| $ with group velocity $v_g$ for frequencies near $\omega_0$), $g(k)=g(-k)$ is the atom-field coupling strength, and we assumed $\hbar=1$. As in \cite{r11}, we consider the single excitation manifold of $\hat{H}$, with the photon field initially in the vacuum state $| \left\{  0 \right\}
 \rangle$. The state vector of  the atom-photon field can be thus expanded as
 \begin{eqnarray}
 | \psi(t) \rangle & = &  \left(  \sum_{m=1}^{2} c_m(t) \exp(-i \omega_0 t) \hat{\sigma}_+^{(m)} + \right. \\
 & + & \left. \int dk c(k,t) \exp [-i \omega(k)t] \hat{a}^{\dag}(k)  \right) | g,g,\{ 0 \} \rangle \nonumber
 \end{eqnarray}
 where $c_{1,2}(t)$ are the amplitude probabilities to find the two atoms in the excited $|e \rangle$ level and $c(k,t)$ is the amplitude probability that one photon has been emitted in the waveguide mode with wave number $k$. Assuming that the spectral coupling $g(k)$ and dispersion curve $\omega(k)$ vary slowly around $k= \pm k_0$, with $\omega(k_0)=\omega_0$, the photon field amplitudes $c(k,t)$ can be eliminated from the dynamics, yielding a set of coupled differential-delayed equations for the excitation atomic amplitudes \cite{r11}
\begin{equation}
 \frac{dc_{1,2}}{dt} =  -\frac{\gamma}{2} c_{1,2}(t)- \frac{\gamma}{2} \exp(i \phi) c_{2,1} (t-d/v_g) \Theta(t-d/v_g) 
 \end{equation}
 where $v_g= (d \omega / dk)_{k_0}$ is the group velocity of waveguide mode at frequency $\omega_0$, $\gamma= (4 \pi / v_g) |g(k_0)|^2$ is the spontaneous emission decay rate of the single quantum emitter, $ \phi=k_0 d$ is the phase delay term, and $\Theta(t)$ is the Heaviside function [$\Theta(t)=1$ for $t>0$, $\Theta(t)=0$ for $t<0$]. For $ \phi$ integer multiple than $2 \pi$, superradiance is observed for the initial coherent symmetric superposition of the atomic state 
 \begin{equation}
|  \psi(0) \rangle =\frac{1}{\sqrt{2}} \left( | g,e, \left\{  0 \right\} \rangle + | e,g, \left\{  0 \right\} \rangle \right).
 \end{equation}
  Equation (4) clearly shows that the spontaneous decay of the two atoms is independent in the time interval $0<t< d/v_g $, with an exponential behavior of the survival probability 
 \begin{equation}
 P_s(t)=|c_1(t)|^2+|c_2(t)|^2
 \end{equation} 
 with decay rate $\gamma$. However, for $t> d/v_g$ retarded interaction between the two atoms, mediated by the photon field in the waveguide, takes place, resulting in a non-exponential behavior of $P_s(t)$. The solution to Eq.(4) is rather cumbersome and can be written, after Laplace transform, as a series of exponential functions with decay rates involving the various branches of the Lambert-W function \cite{r11}. The most striking result is that, for a separation distance $d$ of the order of $l_c/2$, a strong non-Markovian dynamics is observed for $t> d/v_g $, with an instantaneous decay rate that can overcome (up to a factor $\sim 4$) the usual $2 \gamma$ superradiance decay rate. Specifically, let us introduce the dimensionless parameter
 \begin{equation}
 \eta \equiv d \gamma / v_g=d / l_c
 \end{equation}
 that measures the time delay $d/v_g$ in units of the spontaneous decay time $ 1 / \gamma$ of each individual emitter. For $\eta \ll1$, retardation effects can be neglected and collective spontaneous emission with a cooperative decay rate $\simeq 2 \gamma$ is observed, according to usual Dicke superradiance theory. However, as $\eta$ is increased and delay effects become non-negligible, the decay behavior undergoes an abrupt change from 
 an exponential law with decay rate $\gamma$ at $t< d/ v_g $, to a non-exponential law with faster decay at $t>d/v_g $. The largest instantaneous decay rate $\sim 4 \gamma$ is found for  $ \eta \simeq 0.56$ \cite{r11}. 
 \par
 {\it Photon escape dynamics in a waveguide lattice.}  
 Light transport in coupled waveguide structures \cite{r12,r12b,r13,r14,r15,r16,r17,r18,r19,r20,r21,r22,r23} has been harnessed on many occasions to emulate in classical optics a wide variety of hard-to-observe quantum phenomena in the matter \cite{r13}. When light propagates along an optical waveguide $W$ side-coupled to a slab or to a waveguide lattice, it undergoes a nearly exponential decay due to evanescent mode coupling. Such a decay behavior is analogous to the one of a metastable quantum state into a continuum \cite{r13}, like in the spontaneous emission process of an excited atom in a featureless continuum of electromagnetic modes. Non-Markovian effects, Zeno-like dynamics \cite{r16,r26} and slow (algebraic) decay exploting optical tunneling have been observed  in some recent experiments \cite{r21}. Such results suggest us that a system of {\em two distant} waveguides $W_1$ and $W_2$, side-coupled to a waveguide lattice in the geometrical setting of Fig.1(b), can display the typical dynamical features underlying the superradiance paradox. 
 We assume that the modes in all guides have the same propagation constant $\omega_0$, and indicate by $J$ the coupling constants between adjacent waveguides in the lattice and by $\rho \ll J$ the coupling constant between waveguides $W_{1.2}$ and the lattice [see Fig.1(b)]. All waveguides are straight along the axial propagation distance $z$. The Hamiltonian of the photon field propagating along the guiding system reads \cite{r17,r18,r19,r20,r21,r22} $\hat{H}=\hat{H}_A+ \hat{H}_F+\hat{H}_{int}$, where  
 \begin{figure}[htb]
\centerline{\includegraphics[width=8.7cm]{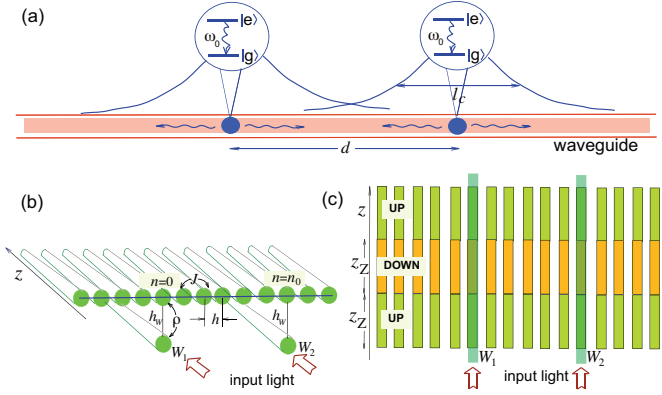}} \caption{ \small
(Color online) (a) Collective spontaneous emission of two identical macroscopically separated two-level atoms (resonance frequency $\omega_0$)  in a one-dimensional waveguide. The waveguide is not chiral and sustains electromagnetic modes with dispersion relation $\omega(k)$ satisfying the time reversal symmetry $\omega(-k)=\omega(k)$, with group velocity $v_g=(d \omega / dk)_{k_0}$ at the frequency $\omega_0=\omega(k_0)$. A superradiance paradox arises when the atom spacing $d$ is in the range $l_c/2<d< l_c$, where $l_c=v_g/ \gamma$ is the coherence length of photons for the single emitter and $\gamma$ the spontaneous decay rate of each individual atom. (b) Waveguide lattice simulation of collective spontaneous emission decay and superradiance paradox. Photon escape dynamics in two dielectric waveguides $W_1$ and $W_2$, side-coupled to a waveguide lattice, emulates the de-excitation dynamics of the two atoms in (a). $\rho$ is the coupling constant between waveguides $W_1$ and $W_2$ with the lattice; $J \gg \rho$ is the coupling constant between adjacent waveguides in the lattice. (c) Zeno dynamics obtained by periodic segmentation of the waveguide lattice at propagation distances multiple than the Zeno distance $z_Z$. The panel shows a view from the top of the two waveguides $W_1$ and $W_2$ side coupled to the waveguide lattice, comprising an alternating sequence of segmented arrays of length $z_Z$ placed at planes up and down the plane of the two waveguides at distances $ \pm h_W$. The interruption of waveguide-lattice evanescent coupling at propagation distances $z=z_Z, 2 z_Z,3 z_Z,...$, emulates a frequent observation of the spontaneous decay process in (a) at time intervals $z_Z$.}
\end{figure} 
 $\hat{H}_A= \omega_0 \sum_{m=1,2} \hat{w}^{\dag}_m \hat{w}_m$ is the Hamiltonian of the photon field in waveguides $W_1$ and $W_2$,  $\hat{H}_F=  \omega_0 \sum_n  \hat{b}^{\dag}_n \hat{b}_n+J \sum_n ( \hat{b}^{\dag}_n \hat{b}_{n+1}+ H.c.)$ is the tight-binding Hamiltonian of the photon field  in the lattice, and $\hat{H}_{int}= \rho (\hat{w}^{\dag}_1 \hat{b}_{0}+\hat{w}^{\dag}_2 \hat{b}_{n_0}+H.c.)$ is the Hamiltonian describing evanescent mode coupling of waveguides $W_1$ and $W_2$ with the lattice. In the above equations, $\hat{w}^{\dag}_{1,2}$ and $\hat{b}^{\dag}_l$ are the bosonic creation operators of photons in waveguides $W_{1,2}$ and in the $l$-th waveguide of the lattice, respectively, $n_0$ is the space distance between the two waveguides, and the space-to-time relation $z=ct$ holds \cite{r12b}  ($z=t$ in units of the speed of light $c$).  
 To describe photon escape dynamics and show the analogy with the spontaneous emission process of two macroscopically-spaced atoms, it is worth switching from Wannier to Bloch basis representation of the photon modes in the lattice \cite{r19,r24} by introducing the bosonic operators $\hat{a}(k) \equiv (1/ \sqrt{2 \pi}) \sum_n \hat{b}_n \exp(-ik n)$ for Bloch modes, where $-\pi \leq k < \pi$ is the Bloch wave number and the bosonic commutation relations 
 $  [ \hat{a}(k), \hat{a}^{\dag} (k^{\prime}) ] = \delta (k-k^{\prime})$  
 hold. In such a representation, the full Hamiltonian of the photon field is given by  $\hat{H}=\hat{H}_A+ \hat{H}_F+\hat{H}_{int}$ with
\begin{eqnarray}
\hat{H}_A & = & \omega_0 ( \hat{w}^{\dag}_1 \hat{w}_1 +  \hat{w}^{\dag}_2 \hat{w}_2)\nonumber \\
\hat{H}_F & = & \int_{-\pi}^{\pi} dk \; \omega(k) \hat{a}^{\dag}(k) \hat{a}(k) \\
\hat{H}_{int} & = & \int_{-\pi}^{\pi} dk \left\{ g(k) \hat{w}^{\dag}_{1} \hat{a}(k) +g(k) \exp(i \phi) \hat{w}^{\dag}_{2} \hat{a}(k) +H.c.  \right\} \nonumber
\end{eqnarray}
where 
\begin{equation}
\omega(k) \equiv \omega_0+ 2 J \cos k
\end{equation}
is the dispersion relation of the tight-binding energy band, $g(k)= \rho / \sqrt{2 \pi}$ is the coupling strength, and $\phi \equiv - n_0 \pi /2$ the phase delay. Note that the Hamiltonian $\hat{H}$ of the bosonic field exactly reproduces the atom-field system Hamiltonian describing collective spontaneous emission provided that the atomic rising operators $\sigma_+^{(1,2)}$ are replaced by the bosonic creation operators $\hat{w}_{1,2}^{\dag}$ in waveguides $W_1$ and $W_2$ [compare Eq.(8) with Eqs.(1) and (2)]. Hence, in the single excitation manifold, photon escape dynamics along the propagation distance $z$ in the integrated photonic structure of Fig.1(b) effectively emulates the collective spontaneous emission process of two macroscopically-spaced atoms of Fig.1(a).  While in the waveguide quantum electrodynamics problem the spontaneous emission decay arises from the atom-field interaction, in our linear optical system photon-photon interaction is prevented and photon decay arises just from evanescent mode coupling (optical tunneling). The distance $d$ is discretized by the integer $n_0$, i.e. $d=n_0$, the resonance condition $\omega(k)=\omega_0$ is attained at $k= \pm k_0= \pm \pi / 2 $, and the group velocity reads $v_g=2J$. As in the collective spontaneous emission problem, we assume the phase delay $\phi=-n_0 \pi /2$  to be an integer multiple than $2 \pi$. The decay rate $\gamma$, given by $\gamma=\rho^2 /J$, corresponds to the independent decay rate of a photon, trapped in either waveguide $W_1$ of $W_2$, into the waveguide lattice owing to evanescent mode coupling. When the time required for the excitation in either one of the two waveguide to reach the other one through the lattice, given by $ n_0/v_g= n_0 / (2 J)$, cannot be neglected and becomes comparable with the photon decay time $ \sim 1/ \gamma$ in the lattice, non-Markovian effects arise in the decay dynamics and the analogue of the superradiance paradox can be observed.  
In the photonic integrated structure, the non-escape probability $P_s(z)$ versus propagation distance $z$, i.e. the analogue of the survival probability of excitation in the collective spontaneous emission problem, can be readily computed from the elements $S_{n,m}(z)$ of the scattering matrix, describing single photon light transport in the structure from input $z=0$ to output $z=z$ planes \cite{r12b,r17,r19,r20,r23}, where $S_{n,m}(z)$ ($n,m=1,2$) is the amplitude probability that one photon injected at $z=0$ in waveguide $W_m$ is found in waveguide $W_n$ at plane $z=z$.
In particular, for the input excitation in the single-photon symmetric superposition state
 \begin{figure}[htb]
\centerline{\includegraphics[width=8.7cm]{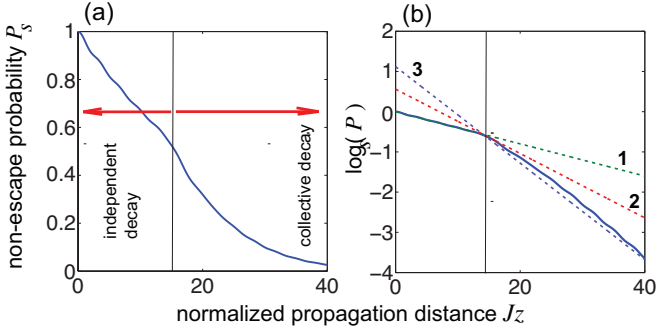}} \caption{ \small
(Color online) (a) Behavior of the non-escape photon probability $P_s(z)$ [Eq.(11)] versus normalized propagation distance $J z$  in the waveguide array of Fig.1(b) for parameter values 
$\rho / J=0.2$ and $n_0=28$. The system is initially excited in the superradiance state defined by Eq.(10). (b) Plot of ${\rm log}(P_s)$ versus $Jz$. The three dashed lines correspond to exponential decay laws with decay rate $\gamma$ (curve 1), $2 \gamma$ (curve 2), and $3 \gamma$ (curve 3), where $\gamma=\rho^2/J=0.04J$ is the photon decay rate for the single waveguide (either $W_1$ or $W_2$) into the lattice. Curve 1 corresponds to independent photon escape at the rate $\gamma$, curve 2 corresponds to collective photon escape at a rate $2 \gamma$ predicted  by Dicke model without delay, and curve 3 corresponds to enhanced superradiance decay.The thin vertical curves in (a) and (b) show the propagation distance $z$ corresponding to the delay $d/v_g=14/J$.} 
\end{figure} 
\begin{equation}
| \psi(0) \rangle= (1/ \sqrt{2})  \left( \hat{w}_1^{\dag}+\hat{w}_2^{\dag} \right) |  0 \rangle,
\end{equation}
 corresponding to a superradiance state, one readily obtains 
 \begin{equation}
 P_s(z)=(1/2)  \left(|S_{1,1}+S_{2,1}|^2+|S_{2,2}+S_{1,2}|^2  \right).
 \end{equation}
 Since we consider single photon propagation in the structure, the non-escape probability $P_s(z)$ can be measured using classical (coherent) input light states (rather than a single photon state), the symmetric input state excitation (10) being readily implemented in an experiment using a balanced optical directional coupler that equally splits the light field into the input waveguides $W_1$ and $W_2$. This makes our suggested photonic simulation of the superradiance paradox of easy experimental implementation, using available integrated photonic technology and classical light excitation. To illustrate non-Markovian features in the photon escape dynamics, Fig.2 shows the behavior of $P_s(z)$ versus propagation distance for parameter values $\rho/J=0.2$ and $n_0=28$, corresponding to a time delay $d/v_g=n_0/(2J)=14/J$ and decay time $1 / \gamma =J/ \rho^2=25 /J$, i.e. to $\eta=0.56$. The behavior of $P_s(z)$
is obtained using Eq.(11), where the scattering amplitudes $S_{n,m}(z)$ have been numerically calculated from the exact Hamiltonian model, i.e. by numerically solving coupled-mode equations of the coupled waveguide system \cite{r12,r13,r20}. The figure clearly shows the transition from independent to collective photon escape into the waveguide lattice, with the onset of strong non-Markovian effects at times $z$ larger than $d/v_g$ and with local decay rates significantly exceeding the superradiance Dicke rate $ 2 \gamma$ [see the slopes in the logarithmic plot of Fig.2(b)].\\ 
\par
{\it Zeno dynamics.}  An interesting feature of the delayed feedback dynamics of Eq.(4) is the possibility to fully cancel any collective behavior in the spontaneous emission process by frequent observations of the system (Zeno dynamics \cite{r25}). Namely, if we  make instant measurements of the system at times $t_Z$, $2 t_Z$, $3t_Z$, ... with $t_Z$ shorter than the time delay $d/v_g$, the collapse of the state after each measurement ensures that the survival probability $P_s(t)$ decays exponentially with the decay rate $\gamma$ of the single quantum emitter. In our integrated photonic setting, instant measurements and state collapse can be implemented by using segmented waveguide lattices \cite{r26,r25b}, as schematically shown in Fig.1(c). In each segmentation interval, light trapped in waveguides $W_1$ and $W_2$ is evanescently coupled either to the upper or to the lower array. Since the arrays are cut, at each propagation lengths $z_Z$, $2 z_Z$, $3z_Z$, ... light escape dynamics restarts with a new array in the vacuum state, thus effectively emulating instant collapse of the wave function (for details see \cite{r26,r25b}).
An example of Zeno dynamics, showing the disappearance of superradiance collective decay, is shown in Fig.3. For typical realistic parameters of waveguide arrays manufactured by femtosecond laser writing in fused silica \cite{
r16,r20,r23,r27}, assuming a coupling constant $J=3.5 \; {\rm cm}^{-1}$ (corresponding to waveguide spacing $h \sim 10 \; \mu$m in the array and $h_W \sim 21 \; \mu$m between waveguides $W_{1,2}$ and the array \cite{r27}), the full length of the array required to observe the dynamics in Figs.2 and 3 is $\simeq 11$ cm, whereas the length of each arrayed segmentation in Fig.1(c) is $z_Z \simeq 2 $ cm.\\
 \begin{figure}[htb]
\centerline{\includegraphics[width=8.7cm]{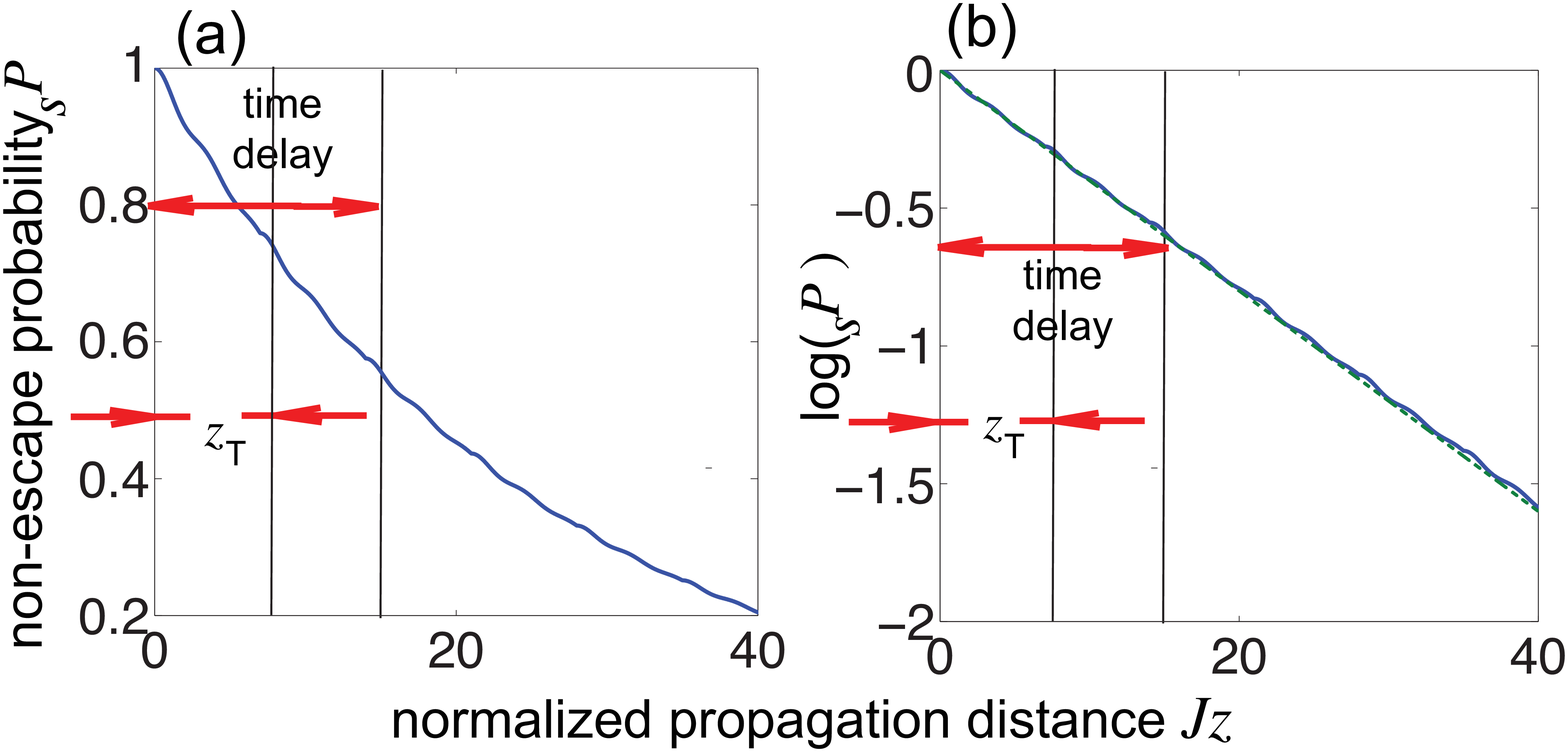}} \caption{ \small
(Color online) Same as Fig.2, but for the segmented array system of Fig.1(c) (Zeno dynamics). The  array segmentation length is $z_T=7/J$, i.e. half than the delay time $d/v_g=14/J$. The dashed line in (b), almost overlapped with the solid one, corresponds to an exact exponential decay law with decay rate $\gamma$, where $\gamma=\rho^2/J=0.04J$ is the photon decay rate for the single waveguide (either $W_1$ or $W_2$) into the lattice.} 
\end{figure}

{\it Conclusions.} Escape dynamics of photons in waveguide lattices can provide a simple and experimentally accessible platform to simulate the spontaneous emission process of two macroscopically separated atoms on a waveguide in the single excitation manifold. Our results suggest a route for the observation of the superradiance paradox in integrated photonic circuits, and could be extended to explore superradiance and non-Markovian effects in case of multiple quantum emitters \cite{r28}.\\
\par

The author declares no conflicts of interest.

\end{document}